\documentclass[11pt,a4paper]{article} 
\usepackage{hyperref} 
\pdfoutput=1 

\begin{document} 
 
\title{Tip streaming from drops flowing in a spiral microchannel } 
 
\author{Simon Molesin and Charles N. Baroud \\ 
\\\vspace{6pt} LadHyX and Department of Mechanics, \\ Ecole
Polytechnique, 91128 Palaiseau, France}

\maketitle 
 
 
\begin{abstract} 
  This fluid dynamics video shows drops of water being transported by a
  mean flow of oil, in a microchannel shaped as a logarithmic
  spiral. The channel shape means that the drops are submitted to an
  increasing shear and elongation as they flow nearer to the center of
  the spiral. A critical point is reached at which a long singular tail
  is observed behind the drops, indicating that the drops are
  accelerating. This is called ``Tip streaming''. 
\end{abstract} 
 
 
\section{Discussion} 

Drops that are submitted to strong shear will reach a critical point at
which they create singular tips. This was first observed by
Taylor~\cite{taylor34} and later studied by de Bruijn~\cite{debruijn93}.

Microfluidics offers the capability to study a whole train of drops,
rather than a single drop at a time. This movie shows how a sequence of
drops reach their critical shear and elongation at which they produce
tip-streaming. See presentation GG.00003 for the physics behind the
tips.

\href{http://hdl.handle.net/1813/11465}{The video can be viewed online 
 by clicking here.}


\end{document}